\documentclass[conference,a4paper]{IEEEtran}
\IEEEoverridecommandlockouts
\ifCLASSINFOpdf

\else
  \fi
\usepackage[pdftex]{graphicx}
\usepackage{multirow}
\usepackage{balance}
\usepackage{here}
\renewcommand\IEEEkeywordsname{Keywords}

\begin{document}
\title{Improving Compression Based Dissimilarity Measure for Music Score Analysis
}

 \author{\IEEEauthorblockN{Ayaka Takamoto\IEEEauthorrefmark{1}, Mayu Umemura\IEEEauthorrefmark{2}, Mitsuo Yoshida\IEEEauthorrefmark{1} and Kyoji Umemura\IEEEauthorrefmark{1}}
 \IEEEauthorblockA{\IEEEauthorrefmark{1}Department of Computer Science and Engineering\\
 Toyohashi University of Technology \\
 Toyohashi, Aichi, Japan\\
 Email: a153350@edu.tut.ac.jp, yoshida@cs.tut.ac.jp, umemura@tut.jp
 }
 \IEEEauthorblockA{\IEEEauthorrefmark{2}Department of Scientific and Engineering Simulation\\
 Nagoya Institute of Technology \\
 Nagoya, Aichi, Japan\\
 Email: umemayu.nitech@gmail.com
}
 }

\maketitle
\IEEEpubid{\makebox[\columnwidth]{978--1--5090--1636--5/16/\$31.00~\copyright~2016 IEEE \hfill }
\hspace{\columnsep}\makebox[\columnwidth]{\hfill }}

\begin{abstract}
In this paper, we propose a way to improve the compression based dissimilarity measure, CDM. 
We propose to use a modified value of the file size, where the original CDM uses an unmodified file size. 
Our application is a music score analysis. 
We have chosen piano pieces from five different composers. 
We have selected 75 famous pieces (15 pieces for each composer). 
We computed the distances among all pieces by using the modified CDM. 
We use the K-nearest neighbor method when we estimate the composer of each piece of music. 
The modified CDM shows improved accuracy. 
The difference is statistically significant. 
\end{abstract}


\section{Introduction}
\IEEEpubidadjcol
We have been interested in the task of classification of music scores based on the patterns in those scores \cite{Dannenberg1997,Sawada2000}. 
The compression based dissimilarity measure, CDM, is proposed for the analysis of time series of data \cite{CDM}. 
The principle of CDM is as follows: the more patterns two strings share, the smaller is the compressed file size of their concatenated string. 
Normalized compression distance (NCD) \cite{Alg,Clustering}, which is uses same principle, is reported to be useful to measure the similarity between polyphonic music scores \cite{Alg, CDM_music} . 
Since one composer may have a favorite pattern within a  score, classification according to the composer seems a more suitable problem for CDM or NCD rather than classifying music scores according to their genre. Actually, it is reported that NCD works for the estimation of the composer of music pieces \cite{Anan}. 
Interestingly, NCD and CDM have been reported to work for estimating the composer of the music pieces \cite{Anan,Adachi2013}. 

We are interested in finding a better way of using compression based similarity measure.
CDM computes similarity based on the existence of a shared pattern in two strings.
As one composer may have a favorite pattern within a score, classification according to the composer seems a more suitable problem for CDM or NCD instead of classification according to genre. 
Therefore, we have selected a composer estimation task for the testbed of CDM.

Usually existing compression programs such as BZIP2, ZIP or GZIP can be used in order to measure the quantity of information of strings \cite{CDM, Alg, Anan}. 
Although the compressed file size should be a reasonable approximation of information quantity of the original string, we may obtain a better approximation with a more careful treatment of the file size. 

In this paper, we verify that the accuracy of the result is highly dependent on the kind of compression program, and that determining a suitable compression algorithm is important for this task. 
Then, we investigate the relationship between the quantity of information and the compressed file size and we explain how to modify the file size in order to obtain better approximation of information quantity from the file size. 
Finally, we show a statistically significant difference by using the obtained approximation. 

\IEEEpubidadjcol
\section{Compression based Dissimilarity Measure}
In this section, we will describe the CDM and its principles. 
Let $x, y$ be some strings that we are interested in. 
Let $xy$ be the string which is the concatenation of string $x$ and $y$. 
Let $C(x)$ be the size of the compressed file, where its original file consists of string $x$. 
CDM is defined as follows:
\[
CDM(x,y)=\frac{C(xy)}{C(x)+C(y)}.
\]
The value of $C(x)$ can be regarded as a practical approximation of the information quantity in string $x$. 
In an extreme case, if $x$ is a repetitive series of an identical pattern, the information quantity of $x$ and the value of $C(x)$ become much smaller than the length of string $x$. 
If two strings are similar, $C(xy)$ becomes smaller than $C(x)+C(y)$ because $C(x)+C(y)$ double counts the information that is shared by string $x$ and string $y$, whereas $C(xy)$ does not. 
Therefore, the value of CDM reflects the amount of shared patterns of the two strings $x$ and $y$.

The actual value of $C(x)$ depends on the kind of compression program. 
Although the value of $C(x)$ should be the information quantity of string $x$ in an ideal case, it is known that there is no compression program that can compress files into a file whose size is equal to the information quantity. 
There are two reasons for this. The first reason is every compression program may not be able to capture certain kinds of patterns. For example, the compression program named ZIP uses the dictionary of string. 
There is maximum length of string in the dictionary because of memory limitation. If the string contains very long patterns, ZIP may not capture the pattern. 
The second reason is that the compressed file may contain extra information that is not related to the information quantity of the input. 
The example of extra information is the first part of the compressed file. 
Usually the first part contains the information about the file format, which is independent from the original file. 
There may be different information in the first part if the compression program used is different. 
In an extreme case, the string whose length is zero contains no information, and its information quantity is also zero. 
Still, the size of the compressed file for this string may not be zero.

The proposed method is first to decide appropriate compression program for this task, and then to obtain more accurate information quantity from compressed file size.

\section{String Representation of Music Scores}
\begin{figure}[t]
\begin{center}
\includegraphics[width=80mm]{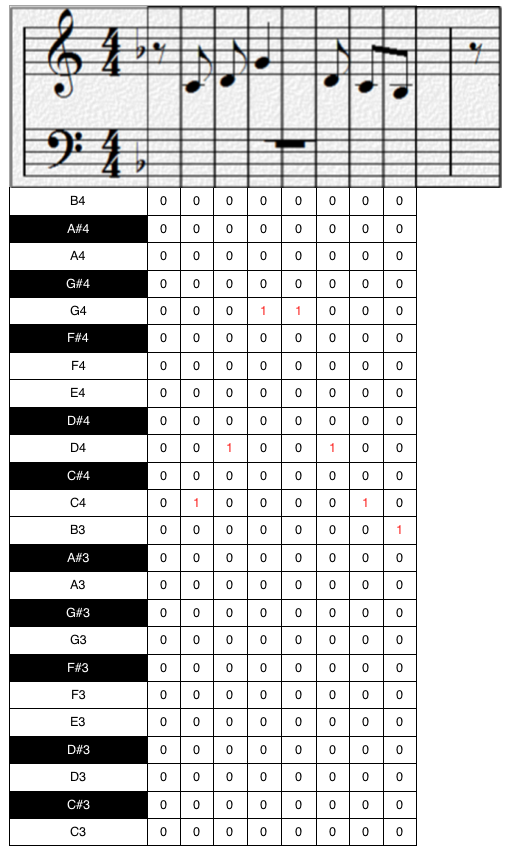}
\caption{Music score and sequence of bit vectors. Each pitch corresponds to one dimension of bit vectors. Since a piano has 88 keys, every bit vector actually used is an 88-dimensional vector. A one-bit vector is obtained at every semiquaver (sixteenth note) interval.}
\label{fig_piano}
\end{center}
\end{figure}

\begin{figure}[t]
\begin{center}
\includegraphics[width=80mm]{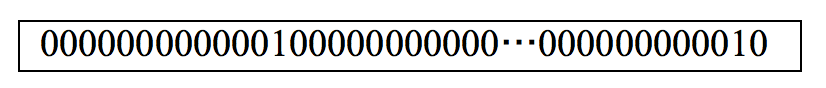}
\caption{A string representation of score in Fig. \ref{fig_piano}. Though it consists of 96 bytes, only the 2 vectors in beginning and the last vector are shown.}
\label{fig_String}
\end{center}
\end{figure}

\begin{figure}[t]
\begin{center}
\includegraphics[width=80mm]{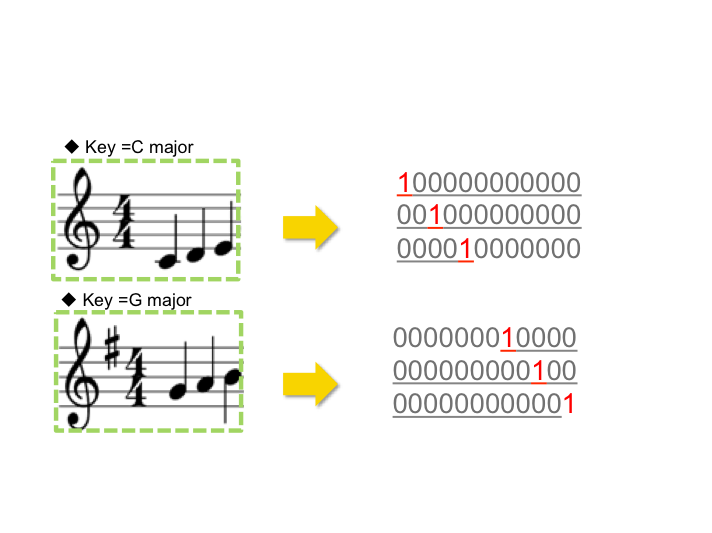}
\caption{Difference in string representation when the score is transposed
The underlined part of string representation is unchanged by a change of key.}
\label{fig_major}
\end{center}
\end{figure}

We need to prepare some string representation of music score before we use CDM. 
We have chosen a simple manner of representing music scores, which has been used in previous research \cite{Adachi2013}. 
First the music score is converted into a sequence of bit vectors. 
Each vector corresponds to the on/off of all the keys of a piano at a certain timing. 
A One-bit vector is obtained at every semiquaver (sixteenth note) interval. 
An example of a sequence of bit vector is illustrated in Fig. \ref{fig_piano}. 
Though the bit vector in Fig. \ref{fig_piano} is 24-dimensional, the actual bit vector is 88-dimentional because there are 88 keys. 
Notes longer than semiquaver are expressed in a series of identical vectors. 
Fig. \ref{fig_String} illustrates the example of string representation from the bit vector in Fig. \ref{fig_piano}. 
In string representation, each bit vector is converted into a sequence of characters using corresponding characters '0' and '1'. Then, the sequences of the characters are concatenated without separators. The result is a very long sequence of  '0' or '1'.  
By this concatenation, even if the music score changes key, almost all parts of the string are unchanged. When the music score changes key, the difference in the string representation is only in the first part and the last part as is illustrated in Fig. \ref{fig_major}.

\section{Estimating Composer}
\begin{figure}[t]
\begin{center}
\includegraphics[width=80mm]{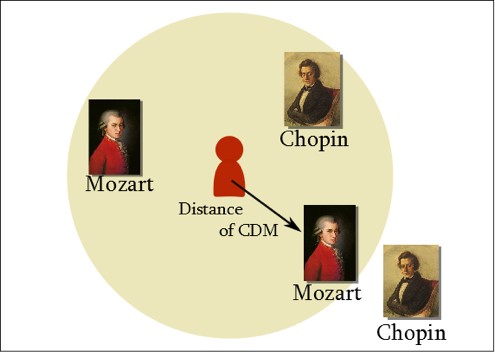}
\caption{Estimating composer, each portrait represents the music score of that composer. 
In this figure, the unknown composer in the middle is estimated as Mozart, since Mozart is a majority within the colored circle.}
\label{fig_composer}
\end{center}
\end{figure}

We estimate the composer of a music score using the K-nearest neighbor method. 
The distance between two music scores is determined by CDM. 
Fig. \ref{fig_composer} illustrates the estimation procedure. 
Intuitively speaking, this procedure is a majority vote. 
We assume that the majority of pieces in the neighbors are by the same composer. 
It is known that the number of neighbors may affect the accuracy of estimation. 
We intuitively set this number to 5 and the estimation works well, using this number. 
We gathered 75 different piano scores. 
Their composers are Bach (Table \ref{Bach}), Mozart (Table \ref{Mozart}), Chopin (Table \ref{Chopin}), Debussy (Table \ref{Debussy}) and Satie (Table \ref{Satie}). 
We selected 15 pieces from each composer. 
Then, we estimated the composer of each piece assuming the composer of the piece is unknown, and we tested whether the result of the K-nearest neighbor estimation is the same as the actual composer. 
The result of each experiment is a set of 75 binary results of correct or incorrect.

\begin{table}[t]
\begin{center}
\caption{Music Titles of Bach}
\label{Bach}
\begin{tabular}{c|p{7cm}}
\hline
Bach&\\
\hline

01 & Chaconne -6 Solo a violino senza basso accompagnato, BWV1004\\
02 & French Suite No.5 in G major,BWV816\\
03 & Air on G String\\
04 & Invention No.1 in C major, BWV772\\
05 & Invention No.8 in F major, BWV779\\
06 & Cantata BWV140, "Wachet auf, ruft uns die Stimme"\\
07 & The Well-Tempered Clavier Book1 Prelude and Fugue No. 2 in C minor, BWV 847 \\
08 & The Well-Tempered Clavier Book1 No.3 in C$\sharp$ major, BWV 848 \\
09 & The Well-Tempered Clavier Book1 No.6 in D minor, BWV 851\\
10 & The Well-Tempered Clavier Book1 Prelude and Fugue No.1 in C major, BWV 846 \\
11 & Invention No.2 in C minor, BWV773\\
12 & Invention No.3 in D major, BWV774\\
13 & Invention No.4 in D minor, BWV775\\
14 & Invention No.5 in E♭minor, BWV776\\
15 & Invention No.6 in E major, BWV777\\
\hline
\end{tabular}
\end{center}
\end{table}

\begin{table}[t]
\begin{center}
\caption{Music Titles of Mozart}
\label{Mozart}
\begin{tabular}{c|p{7cm}}
\hline
Mozart &\\
\hline
01 & Ave verum corpus\\
02 & Agnus Dei\\
03 & Sehnsucht nach dem Fr\"uhlinge K.596\\
04 & Piano Sonata No,11 K.331, 3rd Movement\\
05 & Wiener Sonatinen No.1, 4th Movement\\
06 & Piano Sonata No.8 K.310, 1st Movement\\
07 & Opera“Don Giovanni” Aria di Zerlina\\
08 & Fugue for piano in G K.375g (fragment)\\
09 & Piano Sonata No.15 K.545, 3st Movement\\
10 & Minuet in F major, K.2 \\
11 & Piano Sonata No.11 K.331, 2nd Movement\\
12 & March K.544 (fragment)\\
13 & Piano Sonata No.15 K.545, 2nd Movement\\
14 & Piano Sonata No.3 K.281, 1st Movement\\
15 & The Marriage of Figaro\\
\hline
\end{tabular}
\end{center}
\end{table}

\begin{table}[t]
\begin{center}
\caption{Music Titles of Chopin}
\label{Chopin}
\begin{tabular}{c|p{7cm}}
\hline
Chopin&\\
\hline
01 & Etude Op.10, No.3, in E major\\
02 & Etude Op.10, No.4, in C$\sharp$ minor \\
03 & Etude Op.10, No.5, in G$\flat$ major\\
04 & Etude Op.10, No.12, in C minor ’Revolutionary'\\
05 & Etude Op.25, No.9, in G$\flat$ major\\
06 & Etude Op.25, No.12, in C minor\\
07 & Grande valse brillante, Op.18\\
08 & Polonaise NO.6 Op.53 (Heroic Polonaise)\\
09 & Muzurka Op.7, No.1, Vivace in B$\flat$ major\\
10 & Waltz Op.64, No.1 in D$\flat$major(Minute Waltz)\\
11 & Prelude Op.28 No.15 in D$\flat$ major 'Raindrop'\\
12 & Waltz Op.69, No.1 in A$\flat$ major\\
13 & Muzurka Op.33, No.2, in D major\\
14 & Nocturne Op.9 No.1 in B$\flat$ minor\\
15 & Nocturne Op.9 No.2 in E$\flat$ major\\
\hline
\end{tabular}
\end{center}
\end{table}

\begin{table}[t]
\begin{center}
\caption{Music Titles of Debussy}
\label{Debussy}
\begin{tabular}{c|p{7cm}}
\hline
Debussy&\\
\hline
01 & Preludes Book1, La fille aux cheveux de lin(The Girl with the Flaxen Hair)\\
02 & Arabesque No.1 Andantino con moto, Deux arabesques\\
03 & Arabesque No. 2. Allegretto scherzando, Deux arabesques\\
04 & Beau Soir\\
05 & Preludes Book1, La cath\'edrale engloutie(The Submerged Cathedral)\\
06 & Children's Corner "Golliwogg's Cakewalk"\\
07 & Le petit n\`egre \\
08 & Suite bergamasque "Passepied"\\
09 & Suite bergamasque "Pr\'elude"\\
10 & R\^everie\\
11 & Suite bergamasque "Menuet"\\
12 & Suite bergamasque "Clair de lune"\\
13 & Children's Corner "Doctor Gradus ad Parnassum"\\
14 & Children's Corner "Serenade of the Doll"(Serenade for the Doll)\\
15 & Chileden's Corner "The Snow is Dancing"\\
\hline
\end{tabular}
\end{center}
\end{table}

\begin{table}[t]
\begin{center}
\caption{Music Titles of Satie}
\label{Satie}
\begin{tabular}{c|p{7cm}}
\hline
Satie &\\
\hline
01 & 6 Pieces de la periode 1906~13 "D\'esespoire agr\'eable"\\
02 & Fantaisie - Valse\\
03 & Gnossienne No.1\\
04 & Gnossienne No.2\\
05 & Gnossienne No.3\\
06 & Je te veux\\
07 & Poudre d'or - valse\\
08 & Reverie du pauvre\\
09 & 6 Pieces de la periode 1906~13 "Songe creux"\\
10 & Tendrement\\
11 & Vexations\\
12 & Gymnop\'edie No.1 \\
13 & Gymnop\'edie No.2\\
14 & Gymnop\'edie No.3\\
15 & Ogive No.1\\
\hline
\end{tabular}
\end{center}
\end{table}

\section{Statistical Test}
We apply the McNemar's test when we compare the results of experiments undertaken with different conditions. 
McNemar's test is a standard non-parametric test when we need to compare binary results. 
Suppose we have a binary result xi from the original method (method1) and another result $y_{i}$ from the improved method (method 2). 
We can form the confusion matrix shown in Table \ref{Table_Mtest}. 
Let $b$ be the number of results where $x_{i}$ is correct and $y_{i}$ is incorrect. 
Let $c$ be the number of results where $x_{i}$ is incorrect and $y_{i}$ is correct. 
If we assume that there is no improvement, in other words, that the accuracies of two methods are equal, then the values of $b$ and $c$ behave like the result of a fair coin toss. 
We can reject this hypothesis if $b$ is much smaller than $c$. The p-value of this test is as follows:
\[
p=\left(\frac{1}{2^{b+c}}\right)\sum_{k=0}^{b}C_k \label{2}.
\]
\begin{table}[t]
\begin{center}
\caption{The Confusion Matrix of McNemar’s Test}
\label{Table_Mtest}
\begin{tabular}{c c | c c |c}
\hline
& & \multicolumn{2}{|c|}{Method2} & \\
& & Correct result & Incorrect result & Total \\
\hline
\multirow{2}{*}{Method 1} & Correct result & a & b & a+b \\
 & Incorrect result & c & d & c+d \\
\hline
& Total & a+c & b+d & N \\
\hline
\end{tabular}
\end{center}
\end{table}
\section{Choice of Compression Program}
First, we need to decide which the compression program to use. 
We conducted experiments using three compression programs. They are BZIP2, ZIP, and GZIP. 
We are particularly interested in using BZIP2 \cite{bzip2} because it uses a block-sort algorithm.

Since the string representation of a music score is a sequence of 88 byte units, the length of the pattern in the string may be longer than 100 characters. 
It is known that the block-sort algorithm can compress the file of very long sequence of patterns.

Table \ref{Table_byComp} shows the number of correct results for the methods using different kinds of compression programs. 
Table \ref{Table_pvalue} shows the value of the confusion matrix for the statistical test, and its p-value, comparing BZIP2 with another program.  
The test suggests that BZIP2 is a suitable choice, as we predicted, and $\alpha$ level of the result is 5\% at least.

\begin{table}[t]
\begin{center}
\caption{The Number of Correct Results for Each Compression Program}
\label{Table_byComp}
\begin{tabular}{c|c}
\hline
Method&The number of correct results out of whole result \\
\hline
by bzip2 & 41/75 \\
by gzip & 30/75 \\
by zip & 17/75 \\
\hline
\end{tabular}
\end{center}
\end{table}

\begin{table}[t]
\begin{center}
\caption{Comparison of Compression Program}
\label{Table_pvalue}
\begin{tabular}{c|c c c c c |c}
\hline
Method 1 - Method 2 & a & b & c & d & N & p \\
\hline
gzip - bzip2 & 20 & 10 & 21 & 24 & 75 & 0.03 \\
zip - bzip2 & 11 & 6 & 30 & 28 & 75 & 0.0003 \\
\hline
\end{tabular}
\end{center}
\end{table}

\section{Offsetting the Compressed File Size}
\begin{figure}[h]
\begin{center}
\includegraphics[width=80mm]{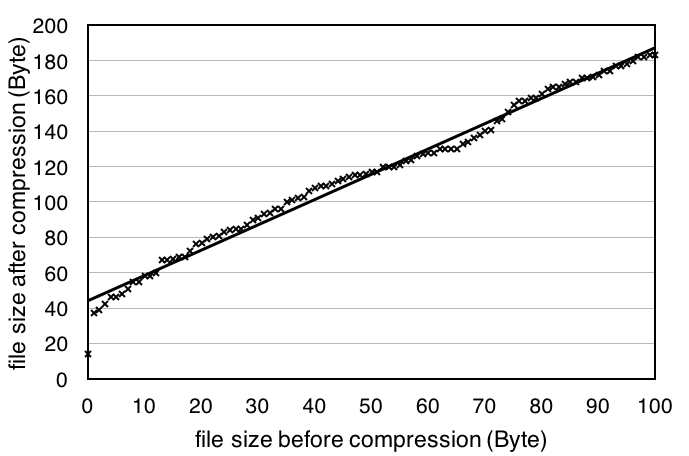}
\caption{The compressed file size vs. information quantity of input. This plot suggests that bzip2 has additional 45 bytes, which is independent from the information quantity of input.}
\label{fig_fileSize}
\end{center}
\end{figure}

There is a straightforward way to check whether the compressed file contains the information that is independent from the input file. 
First, we prepare the files whose length is from 0 to 100, and whose content is a random sequence of characters. 
We cannot compress a random sequence, and the length of the input file reflects the information quantity of the input. 
Then, we compressed each file and obtained the result. 
The result is shown in Fig. \ref{fig_fileSize}, and we draw a line from the points by regression. 
From this line, we may say that there are an additional 45 bytes for each compressed file. 
Therefore, we may estimate the information quantity of input as $(C(x) - 45)/$ $\beta$，where $\beta$ is some positive constant, when we use BZIP2. 
We use $(C(x) - 45)/$ $\beta$ instead of $C(x)$. Please note that the constant $\beta$ does not affect the value of CDM, and therefore we need not to obtain the value of $\beta$ from the line. 
We call the method using this estimated value as CDM with offset.

\section{Effect of offset}
The confusion matrix CDM without offset and CDM with offset is shown in Table \ref{Table_offset_result}
The precision of CDM with offset is 48/75, whereas the precision of CDM without offset is 41/75. It shows some improvements. 
The p-value of McNemar's test becomes 0.02 as shown in equation \ref{eq_pvalue}. 
The difference is statistically significant with $\alpha = 0.05$.
\begin{eqnarray}
p &=& {}_{9}C_{1}(0.5)^8(0.5)^1+{}_{9}C_{0}(0.5)^9(0.5)^0 \nonumber \\
&=& 0.01953\ldots=0.02 \label{eq_pvalue}.
\end{eqnarray}

It should be noted that b is only 1. 
This means that CDM with offset shows a consistent improvement.

\section{Verification of Offset Value}
It is an interesting question as to whether the offset value obtained from compressing random sequences, is actually the best value for CDM. 
To answer this question, we conducted experiments changing the offset value from 0 to 100. 
The result is shown in Table \ref{Table_offset_effect}. 
The result shows that we can get the highest level of precision when the offset is 45. 
This suggests that the improvement in precision is the result of the offset from file size to information quantity.

\section{Discussion}
The experiment of the musical analysis shows that BZIP2 is suitable for CDM. 
When we do not have knowledge about the length of the pattern to capture, it may be the same situation with musical analysis. 
Therefore, BZIP2 is recommended in this case.

The offset values of compression programs will be different among the compression programs. 
The values may be different even among different versions of the compression program. 
Therefore, the value of offset may not always be 45.

There may be some reason why the proposed offset has not been attracting our attention. 
If the string is long, it becomes difficult to detect the improvement by the offset because the difference in CDM is small. 
Nevertheless, even for a long string, there is no reason to not use the offset of compression file size when we use CDM.

Our experiment uses music scores instead of sounds. Since humans can usually estimate the name of a composer on listening to the sound of music, “sounds" becomes the natural target for the next step in research.  
Style classification can be regarded as a subtask of composer estimation. 
Although style classification based on sound is successful \cite{Tonal}, we are estimating that handling sounds is more difficult than handling scores because the former requires signal-processing techniques.

We do admit that it is hard for us to form theories from the results of our experiment. 
There are, however, some works to form the theories of the composer classification task \cite{Classification,Composer_classification}. 
We are interested in interpreting the results of our experiment from these theories, since compression algorithms are currently attracting the attention of the computer music researchers \cite{music_analysis}.
 
\section{Conclusion}
In this paper, we conducted the composer estimation using CDM. 
Firstly, we have shown that BZIP2 is a suitable compression program and explained the reason in terms of the length of pattern. 
Then, we have shown one practical method to obtain the offset of the compressed file size. 
Finally, we have shown that the obtained offset actually improves the accuracy of estimation. 
\balance

\begin{table}[H]
\begin{center}
\caption{The Number of Correct Estimation by The Presence or Absence of The Offset}
\label{Table_offset_result}
\begin{tabular}{c c|c c | c }
\hline
& & \multicolumn{2}{|c|}{With offset} & \\
& & Correct result & Incorrect result & Total \\
\hline
Without & Correct result & 40 & 1 & 41 \\
 offset & Incorrect result & 8 & 26 & 34 \\
\hline
& Total & 48 & 27 & 75 \\
\hline
\end{tabular}
\end{center}
\end{table}

\begin{table}[H]
\begin{center}
\caption{Effect of Changing The Value of Offset}
\label{Table_offset_effect}
\begin{tabular}{c|c}
\hline
Value of offset & The number of correct results/number of all results \\
\hline
0 & 41/75 \\
20 & 45/75 \\
40 & 47/75 \\
60 & 42/75 \\
80 & 46/75 \\
100 & 25/75 \\
\hline
45&48/75 \\
\hline
\end{tabular}
\end{center}
\end{table}

\bibliographystyle{IEEEtran}
\bibliography{icaicta_takamoto}
\end{document}